\documentclass[twocolumn,prl,aps,showpacs,floatfix,superscriptaddress]{revtex4}
\usepackage{amsmath,amssymb,graphicx}
\usepackage{color}

\begin{document}

\newcommand{\todo}[1]{\textbf{\textsc{\textcolor{red}{(TODO: #1)}}}}
\newcommand{\fcs}{Fe$_{1-x}$Co$_{x}$Si}
\newcommand{\mfs}{Mn$_{1-x}$Fe$_{x}$Si}
\newcommand{\mcs}{Mn$_{1-x}$Co$_{x}$Si}
\newcommand{\cso}{Cu$_{2}$OSeO$_{3}$}

\newcommand{\rxx}{$\rho_{xx}$}
\newcommand{\rxy}{$\rho_{xy}$}
\newcommand{\rxytop}{$\rho_{\rm xy}^{\rm top}$}
\newcommand{\Drxyt}{$\Delta\rho_{\rm xy}^{\rm top}$}
\newcommand{\Sxy}{$\sigma_{xy}$}
\newcommand{\Sxya}{$\sigma_{xy}^A$}

\newcommand{\bco}{$B_{\rm c1}$}
\newcommand{\bct}{$B_{\rm c2}$}
\newcommand{\bao}{$B_{\rm A1}$}
\newcommand{\bat}{$B_{\rm A2}$}
\newcommand{\beff}{$B^{\rm eff}$}

\newcommand{\btr}{$B^{\rm tr}$}

\newcommand{\tc}{$T_{\rm c}$}
\newcommand{\ttr}{$T_{\rm tr}$}

\newcommand{\mb}{$\mu_0\,M/B$}
\newcommand{\dmdb}{$\mu_0\,\mathrm{d}M/\mathrm{d}B$}
\newcommand{\ddmddb}{$\mathrm{\mu_0\Delta}M/\mathrm{\Delta}B$}
\newcommand{\cm}{$\chi_{\rm M}$}
\newcommand{\cac}{$\chi_{\rm ac}$}
\newcommand{\rechi}{${\rm Re}\,\chi_{\rm ac}$}
\newcommand{\imchi}{${\rm Im}\,\chi_{\rm ac}$}

\newcommand{\ozz}{$\langle100\rangle$}
\newcommand{\ooz}{$\langle110\rangle$}
\newcommand{\ooo}{$\langle111\rangle$}
\newcommand{\too}{$\langle211\rangle$}




\title{Specific heat of the skyrmion lattice phase and field-induced tricritical point in MnSi}

\author{A. Bauer}
\affiliation{Physik-Department, Technische Universit\"at M\"unchen,
James-Franck-Stra{\ss}e, D-85748 Garching, Germany}

\author{M. Garst}
\affiliation{Institute for Theoretical Physics, Universit\"at zu K\"oln, 
Z\"ulpicher Str. 77, D-50937 K\"oln, Germany}

\author{C. Pfleiderer}
\affiliation{Physik-Department, Technische Universit\"at M\"unchen,
James-Franck-Stra{\ss}e, D-85748 Garching, Germany}

\date{\today}

\begin{abstract}
We report high-precision measurements of the temperature and magnetic field dependence of the specific heat, $C(T,H)$, across the magnetic phase diagram of MnSi. Clear anomalies establish the skyrmion lattice unambiguously as a thermodynamic phase. The evolution of the specific heat anomalies, the field dependence of the entropy released at the phase transitions, and the temperature versus field dependence of crossover lines provide striking evidence of a tricritical point at $\mu_0 H^{\rm int}_{\rm TCP}=340\,{\rm mT}$ and $T_{\rm TCP}=28.5\,{\rm K}$. The existence of this tricritical point represents strong support of a helimagnetic Brazovskii transition, i.e., a fluctuation-induced first order transition at $H=0$.
\end{abstract}


\vskip2pc

\maketitle


In the B20 transition metal compound MnSi the paramagnetic to helimagnetic transition and the discovery of a skyrmion lattice have attracted great scientific interest. A comprehensive neutron scattering study recently provided strong evidence that the helimagnetic transition is driven first order \cite{Matsunaga:JPSJ1982,Pfleiderer:JMMM2001,Stishov:PRB2007} by the interactions between isotropic critical fluctuations \cite{Janoschek:preprint2012}, as predicted theoretically by Brazovskii \cite{Brazovskii:JETP1975}. However, the weak first order character of the helimagnetic transition has also been attributed to the formation of a skyrmion liquid that preempts generically the transition to long range magnetic order in chiral helimagnets \cite{Roessler:Nature2006,Pappas:PRL2009,Hamann:PRL2011}. A stringent test of the Brazovskii-scenario concerns the influence of a magnetic field which is predicted to quench the interactions between the critical fluctuations, thereby changing the transition from first to second order at a tricritical point. 

The skyrmion lattice consists of a regular arrangement of topologically non-trivial whirls in the magnetic texture. Its formation has been explained in terms of a single thermodynamic phase stabilized by thermal Gaussian fluctuations \cite{Muehlbauer:Science2009,Adams:PRL2011, Tonomura:NanoLetters2012}. However, a competing theoretical study predicted a complex generic phase diagram of B20 helimagnets supporting various topological textures and meso-phases, driven by a combination of magnetic anisotropies and a softened modulus of the magnetization \cite{Butenko:PRB2010}. Putative experimental evidence of this complex phase diagram was inferred for FeGe, another B20 compound, from measurements of the ac susceptibility \cite{Wilhelm:PRL2011} and essentially featureless specific heat data \cite{Wilhelm:JPCM2012}. The experimental evidence for the complex phase diagram was questioned by comprehensive simultaneous measurements of the ac susceptibility and magnetization in MnSi, clearly suggesting the formation of a single skyrmion lattice phase \cite{Bauer:PRB2012}. The perhaps most important outstanding experimental challenge concerns thereby evidence for clear specific heat anomalies as proof of the skyrmion lattice as a thermodynamic phase.


The precision and density of the specific heat data required to resolve these questions does not permit to use a conventional heat pulse technique for at least two reasons. First, typical pulse sizes are of the order of percent of the sample temperature. However, to study the questions of interest here pulses of at most a few tenths of a percent are acceptable to prevent averaging out details. Second, the time required for a single data point of at least a few minutes renders measurements at very finely spaced temperatures and fields unfeasible. 

To obtain high-resolution data for a large number of magnetic field values across the entire phase diagram we used a quasi-adiabatic large pulse technique on a Quantum Design physical properties measurement system (PPMS). This method represents essentially a type of adiabatic scanning calorimetry \cite{Junod:JPSI1979}. Typical heat pulses had a size of 30\% of the sample temperature. The temperature was logged as a function of time both while heating at a constant power and while cooling after switching off the heater. The specific heat reported here was inferred from the cooling curves where we applied a moving average over 3\% of the data points in a relaxation curve. The first 1\% of every pulse was excluded from the analysis. The results obtained are perfectly consistent with data reported in Ref.\,\cite{Bauer:PRB2010} where a conventional small heat pulse method with pulses of 1\% to 2\% was used.  Control measurements with tiny heat pulses of 0.2\,\% confirmed that the internal time constants of the sample were much smaller than the changes of sample temperature. Using this method data at 86 different field values were recorded starting from two or three different temperatures at each field to confirm the results. It is interesting to note that  temperature scans with a comparable resolution using a conventional heat pulse technique would require nearly one day per magnetic field.

\begin{figure*}
\includegraphics[width=1.0\linewidth,clip=]{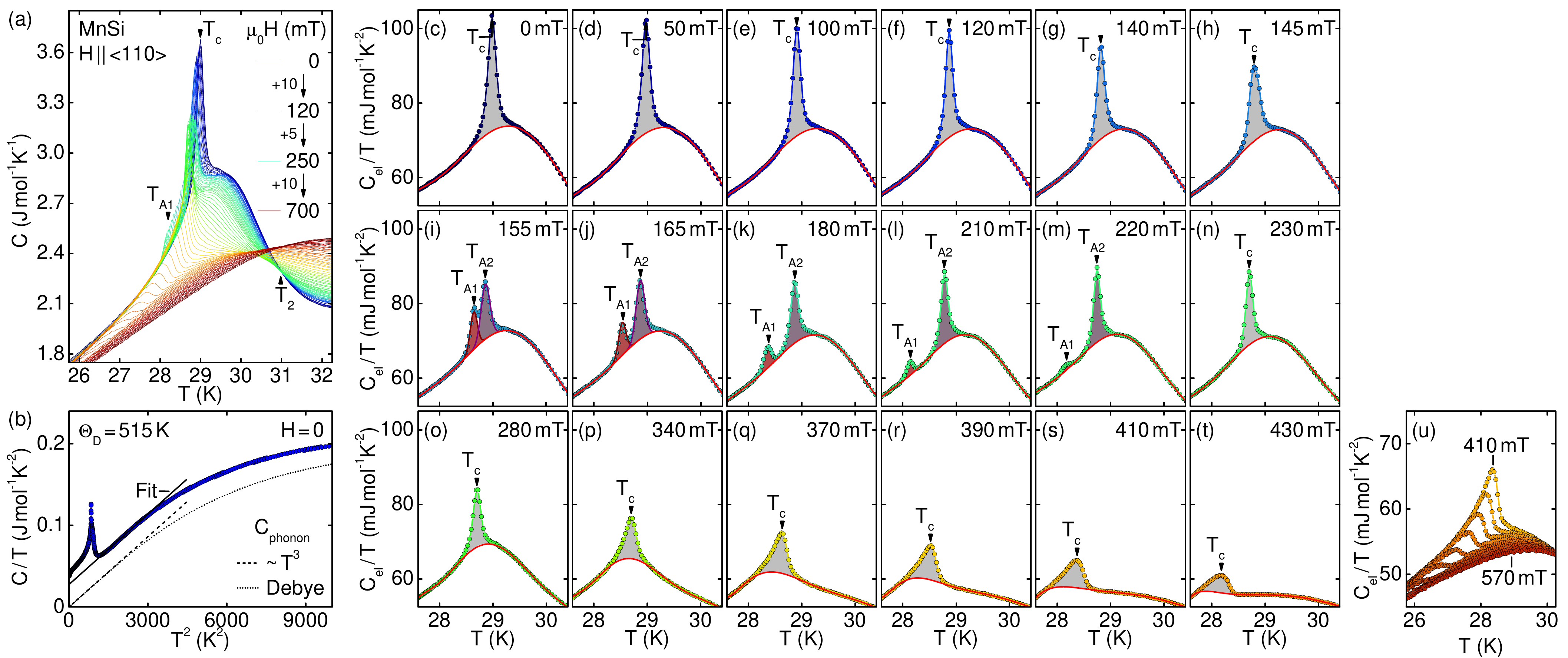}
\caption{(Color online) Specific heat of MnSi as a function of temperature for magnetic fields up to 0.7\,T along the $\langle110\rangle$ axis. At $H=0$ the narrow peak at the helimagnetic transition, marked $T_{c}$, resides on top of a broad shoulder. In small fields the shoulder is characterized by a Vollhardt invariance at $T_2$. In the field range of the skyrmion lattice the peak at $T_{c}$ splits into two peaks where we mark the lower temperature peak as $T_{A1}$. (a) Data as measured. (b) $C/T$ versus $T^2$ to illustrate the lattice subtraction. Panels (c) through (u): specific heat over temperature after subtraction of lattice contributions. Solid lines represent polynomial estimates of the background at temperatures far away from the anomalies of the transition.
}
\label{Figure01}
\end{figure*}

\begin{figure}
\includegraphics[width=1.0\linewidth,clip=]{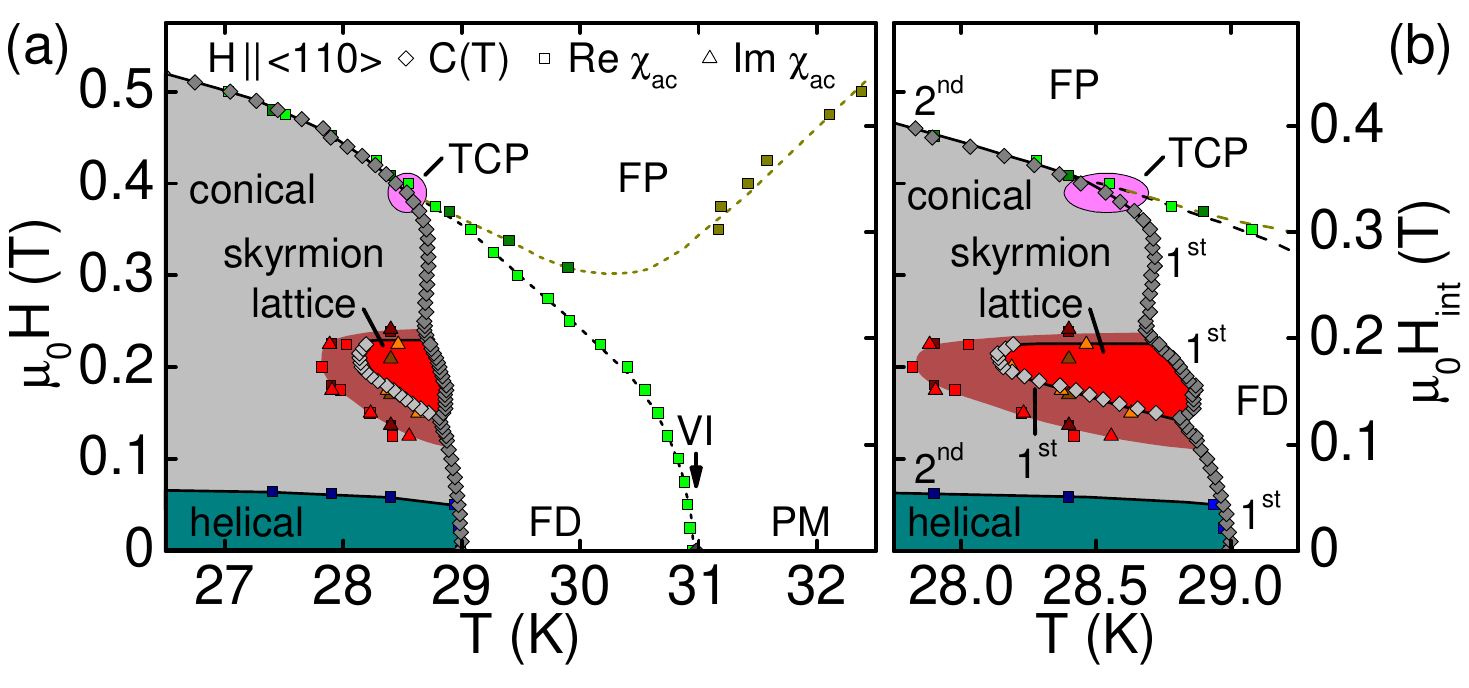}
\caption{(Color online) Magnetic phase diagram of MnSi for field along {\ooz} as derived from the specific heat (diamonds), as well as the real part (squares) and imaginary part (triangles) of the ac susceptibility, cf.\,Ref.\,\cite{Bauer:PRB2012}. We distinguish the following regimes: helical  order, conical order, skyrmion lattice, fluctuation-disordered (FD), paramagnetic (PM), and field-polarized (FP). Phase transitions and crossover lines are marked as solid and dashed lines, respectively. In the shaded coexistence regime between the conical and skyrmion lattice phase a finite Im$\chi(\omega)$ is observed, characteristic of hysteresis associated with a first order transition. Panel (b) shows an enlarged section of panel (a).
}
\label{Figure02}
\end{figure}

For our study a large single crystal was grown by optical float-zoning under ultra-high vacuum compatible conditions (for details see Refs.\,\cite{Neubauer:RSI2011,Bauer:PRB2010}). The single crystal was oriented by means of Laue x-ray diffraction and samples were cut with a wire saw. The residual resistivity ratio of samples from the same ingot was 70, characteristic of the sample quality studied in most previous investigations. The specific heat was measured on a sample with the shape of a parallelepiped of roughly $4\times1.5\times1\,{\rm mm^{3}}$, for magnetic field parallel to the short side corresponding to a {\ooz} axis (the soft and hard magnetic axes of MnSi are {\ooo} and {\ozz}, respectively). For corrections of demagnetizing fields (cf. Fig.~\ref{Figure02}) the sample was approximated as a cuboid \cite{Aharoni:JAP1998}.

To support the specific heat data we also measured the ac susceptibility of the same sample for the same orientation with a PPMS at an excitation frequency of 911\,Hz and an excitation amplitude of 1\,mT. This way we avoided systematic differences when comparing the magnetic phase diagrams inferred from the specific heat and ac susceptibility, respectively, e.g., due to small differences of demagnetizing fields. The ac susceptibility data was in excellent agreement with previous work \cite{Bauer:PRB2012}. However, to account for a small systematic offset between the sample thermometers the temperature scale of the susceptibility was shifted by $-0.1\,{\rm K}$. 


Shown in Fig.~\ref{Figure01}\,(a) are typical specific heat data of MnSi in the vicinity of the helimagnetic phase transition as a function of temperature for a wide range of magnetic fields. Our data are in excellent agreement with previous reports for the field values available  \cite{Bauer:PRB2010}. For zero magnetic field a sharp peak at the onset of helimagnetic order, marked as $T_{c}$, resides on top of a broad shoulder. For small fields $H$, the latter is characterized by a Vollhardt invariance \cite{Vollhardt:PRL1997} at a temperature $T_2$  where $\partial C /\partial H|_{H = 0} = 0$ \cite{Bauer:PRB2012,Janoschek:preprint2012}. With increasing field a second peak emerges in the range of the skyrmion lattice, marked as $T_{\rm A1}$ in Fig.~\ref{Figure01}. This second peak had not been reported before. For even larger fields the peak at {\tc} shifts to lower temperatures and changes its character from first to second order. Finally, we did not observe evidence suggesting the formation of meso-phases or other complexities anywhere. 

The phonon contribution to the specific heat may be inferred from the cubic temperature dependence of $C(T)$ in the range $T_{2} < T < 50\,{\rm K}$ (Fig.~\ref{Figure01}\,(b)). The corresponding Debye temperature, $\Theta_{\text{D}} = 515\,{\rm K}$, is in excellent agreement with previous work \cite{Bauer:PRB2010} \cite{debye-note}. Subtracting this lattice contribution The detailed field dependence of the electronic contribution to the specific heat divided by temperature, $C_{\rm el}/T$, close to the helimagnetic transition is illustrated in the remaining panels of Fig.~\ref{Figure01}. 

In small applied fields the shape of the peak at {\tc} is well described as a \textit{symmetric} Gaussian with typical values of the full-width at half-maximum of $\Delta T_{\rm c}\sim(0.15\pm0.02)\,{\rm K}$. This is characteristic of the latent heat of a slightly broadened first order transition where we attribute the broadening to imperfections of the sample. The peak resides as an additional feature on a broad shoulder (Figs.~\ref{Figure01}(c) through (h)). Up to 140\,mT the peak at {\tc} is essentially unchanged, while its position shifts slightly to lower temperatures. 

Around 145\,mT a second peak appears at a temperature $T_{\rm A1}$. This peak clearly shifts to lower temperatures with increasing fields and decreases in size until it vanishes above $\sim230\,{\rm mT}$ (Figs.~\ref{Figure01}\,(i) through (n)). In the field range for which two peaks are observed we denote the maximum of the peak at the higher temperature as $T_{\rm A2}$. The shape and size of the peak at $T_{\rm A2}$ as well as the values of $T_{\rm A2}$ are essentially unchanged in this field range. The broadening around 145\,mT is due to the initial overlap of the two peaks which are well described at all fields by Gaussians of widths similar to $\Delta T_{\rm c}$.  

Finally, for field values exceeding the regime of the skyrmion lattice, the shape of the specific heat anomaly changes drastically as in Figs.~\ref{Figure01}\,(o) through (u) (here the maximum is again denoted as {\tc}). For a field around 400\,mT the anomaly is clearly \textit{asymmetric} characteristic of a mean-field lambda anomaly. At the same time the broad shoulder on which the peak resides moves to lower temperatures. For even larger fields the lambda anomaly displays some rounding as shown in Fig.~\ref{Figure01}\,(u) which may be attributed to the field dependence of {\tc}, i.e., temperature scans cut the phase boundary under a shallow angle.


Shown in Fig.~\ref{Figure02}\,(a) is the phase diagram as inferred from our specific heat and susceptibility data. For small magnetic fields, the specific heat displays a crossover with decreasing temperature from an uncorrelated paramagnetic (PM) to a fluctuation disordered (FD) regime. The FD regime is characterized by strongly interacting chiral fluctuations that eventually induce a first-order Brazovskii transition to the magnetically ordered state \cite{Janoschek:preprint2012}. This interaction suppresses the correlation length and thus induces a point of inflection in the susceptibility, $\partial^2 \chi/\partial T^2=0$. Its position defines the crossover line between the PM and FD regimes which meets the Vollhardt invariance (VI) for $H\to0$. At high magnetic fields, where the interaction between the chiral paramagnons is quenched, the transition to the conical phase is of second order and in the XY universality class. Consequently, there must exist a fluctuation-induced tricritical point (TCP) \cite{tricritical} at intermediate fields where the transition changes from first to second order. From general scaling arguments, one expects the crossover between the PM and FD regime to merge with the phase boundary at this singular TCP. Note that also the crossover line between the PM and the field-polarized (FP) regime observed in the susceptibility emanates from the TCP. Taken together, this allows to estimate $T_{\rm TCP}\approx28.5\,{\rm K}$ and $\mu_0 H_{\rm TCP}\approx 390\,{\rm mT}$. The ordinate on the right hand side reflects thereby an estimate of the internal field, $H_{\rm int} = H - N M \approx H (1- N \chi_{\rm con})$, where $\chi_{\rm con}$ is the constant susceptibility within the conical phase. At the TCP we find $\mu_0 H^{\rm int}_{\rm TCP}\approx340\,{\rm mT}$. Around the TCP the shape of the specific heat anomaly changes from being a \textit{symmetric} Gaussian to a mean-field like \textit{asymmetric} lambda anomaly. A detailed analysis of the corresponding entropies to be presented below confirms the position of the TCP (see also supplementary information at \cite{SOM}).

Decreasing the temperature further within the FD regime a first-order transition either to helimagnetic, conical, or the skyrmion lattice phase is observed depending on the applied field. In the field range of the skyrmion lattice, two first-order anomalies at $T_{\rm A1}$ and $T_{\rm A2}$ with $T_{\rm A1} < T_{\rm A2}$ are found as a function of temperature. This establishes the skyrmion lattice unambiguously as a thermodynamic phase. Interestingly, the dissipative part of the ac susceptibility, Im$\chi(\omega)$, is quite different at these two transitions \cite{Bauer:PRB2012}. A comparatively broad coexistence regime with a substantial Im$\chi(\omega)$ encompasses the $T_{\rm A1}$ line (shaded regime between the conical and skyrmion lattice phase in Fig.~\ref{Figure02}\,(b)), whereas at the fluctuation-induced first-order transition at $T_{\rm A2}$ no significant contribution to Im$\chi(\omega)$ is observed. The finite dissipation derives here from the nucleation process of topologically non-trivial skyrmions within the conical phase \cite{Milde:preprint2013}, which eventually triggers the first order transition at $T_{\rm A1}$.


\begin{figure}
\includegraphics[width=1.0\linewidth,clip=]{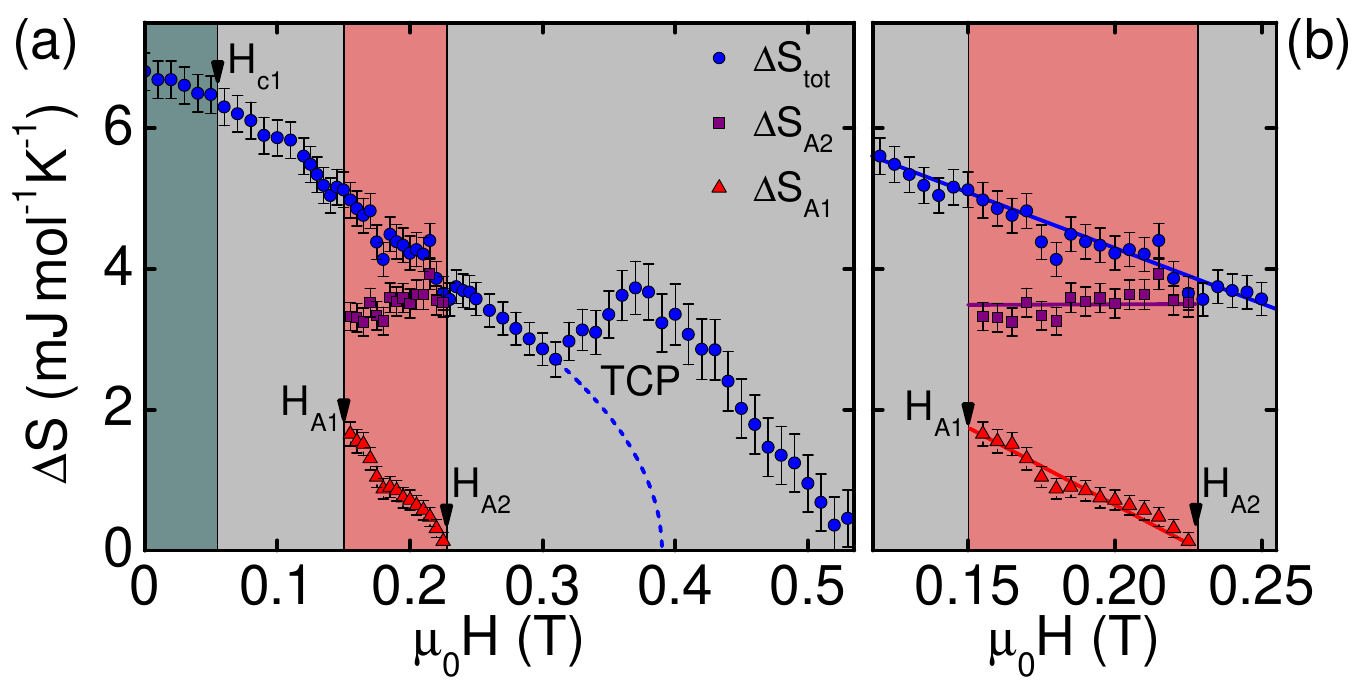}
\caption{(Color online) Entropy, $\Delta S_{\text{tot}}$, inferred from the specific heat anomalies at {\tc}, $T_{\rm A1}$, and $T_{\rm A2}$. Error bars reflect conservative estimates of uncertainties of the background subtraction. (a) For small fields $\Delta S_{\text{tot}}$ essentially corresponds to the latent heat decreasing linearly. Near the TCP critical fluctuations contribute to $\Delta S_{\text{tot}}$ resulting in a local maximum. The dashed line indicates an estimate of the latent heat that vanishes at $\mu_0 H_{\rm TCP}\sim390\,{\rm mT}$. (b) Expanded view of the regime of the skyrmion lattice. 
}
\label{Figure03}
\end{figure}

Further, considering the relationship of Clausius-Clapeyron, $dT/d(\mu_0 H)=-\Delta S/\Delta M$, the negative slope of $T_{\rm A1}(H)$ at the lower boundary of the skyrmion lattice phase implies an increase of entropy as $\Delta M>0$ \cite{Bauer:PRB2012}. In contrast, there is no difference of entropy when exiting the skyrmion lattice phase at larger fields, as the slope is vanishingly small. Therefore, the skyrmion lattice phase yields a larger entropy than the conical phase. This conjecture is supported by the detailed phase boundary between the ordered phases and the fluctuation disordered regime shown in Fig.~\ref{Figure02}\,(b), where $T_{\rm A2}$ is shifted towards higher temperatures as expected of a thermodynamically ordered state with a larger entropy than the (competing) conical phase. 

After subtracting a background to $C_{\rm el}/T$ determined by a polynomial fit (red lines in Fig.~\ref{Figure02}) the shaded areas were integrated to determine the entropy released at the phase transitions, $\Delta S$. The resulting field dependence of $\Delta S$ is summarized in Fig.~\ref{Figure03}\,(a) where Fig.~\ref{Figure03}\,(b) shows the field range of the skyrmion lattice in further detail. The error bars are estimated by a systematic variation of the background contribution; the conclusions discussed in the following are not sensitive to the precise choice of background (see also the supplementary information at \cite{SOM}). For fields exceeding $\sim$300\,mT the anomaly becomes broader and the absolute size of $\Delta S$ depends more sensitively on the choice of the background but the  general evolution of $\Delta S$ remains unaffected. We denote the entropy released near {\tc} by $\Delta S_{\text{tot}}$. In the field range of the skyrmion lattice the entropies of the anomalies at $T_{\rm A1}$ and $T_{\rm A2}$ are denoted as $\Delta S_{\rm A1}$ and $\Delta S_{\rm A2}$, respectively, where $\Delta S_{\text{tot}}$ is determined as the sum of $\Delta S_{\rm A1}$ and $\Delta S_{\rm A2}$. Note that $\Delta S$ is basically the latent heat for the transitions at small fields, whereas $\Delta S_{\text{tot}}$ corresponds to the entropy released near the second-order transition for $H \gtrsim H_{\rm TCP}$.

The latent heat determined this way is in quantitative agreement with Clausius-Clapeyron, suggesting that our data capture $C(T)$ without loss of information. On the one hand, we find for the lower boundary of the skyrmion lattice phase $dT/d(\mu_0 H) = - (12\pm1)\,{\rm K/T}$, corresponding to $dT/d(\mu_0 H_{\rm int}) = - (14\pm1)\,{\rm K/T}$ after correcting demagnetizing fields. On the other hand we observe around the low field and high temperature boundary of the skyrmion lattice phase a change of entropy $\Delta S_{\rm A1} = (1.5\pm0.2)\,{\rm mJ\,mol^{-1}K^{-1}}$. From magnetization measurements we estimate for the change at the first-order transition $\Delta M = (3.4\pm1)\cdot10^{-3}\,{\rm \mu_B\,f.u.^{-1}} = (1.3\pm0.4)\,{\rm kA/m}$. This implies a value of $-\Delta M/\Delta S_{\rm A1} = -(13\pm4)\,{\rm K/T}$. In comparison, at the upper field boundary $\Delta M$ is similar to the lower boundary while $\Delta S_{\rm A1}\to 0$, consistent with the absence of a noticeable slope.

The field dependence of the entropy released at the phase boundaries shown in Fig.~\ref{Figure03} yields several important pieces of information. For small fields, $\Delta S_{\text{tot}}$ can be identified with the latent heat of the first-order transitions and it decreases monotonically from its zero-field value, $6.8\,{\rm mJ\,mol^{-1}K^{-1}}$, across all phases up to $\sim300\,{\rm mT}$. Second, the unchanged monotonic decrease across the skyrmion lattice phase implies that its entropy is larger than that of the conical phase as conjectured above. Moreover, the field dependence originates from the transition between the skyrmion lattice and the conical state at $T_{\rm A1}$, while $\Delta S_{\rm A2}$ is essentially unchanged.  As $\Delta S_{\rm A1}$ vanishes with increasing field the skyrmion lattice phase becomes thermodynamically unfavourable above $H_{\rm A2}$. Third, near the TCP critical fluctuations start to contribute to $\Delta S_{\text{tot}}$ resulting first in an increase as a function of $H$. In a mean-field approximation, the latent heat is expected to vanish as $\Delta S_{\rm latent\, heat} \sim \sqrt{H_{\rm TCP}-H}$ for $H \to H_{\rm TCP}$ (dashed line in Fig.~\ref{Figure03}(a)) which is consistent with the value of $H_{\rm TCP}$ estimated with the help of the crossover lines. For larger fields $H > H_{\rm TCP}$, the fluctuations become less singular and $\Delta S_{\text{tot}}$ again decreases until it is finally expected to vanish as the conical phase is suppressed by magnetic field.
 
In conclusion, we find clear evidence that the skyrmion lattice in MnSi represents a thermodynamic phase without signs of further complexities such as mesophases in high-precision measurements of the specific heat. Moreover, we identify the existence of a tricritical point as a stringent test of the helimagnetic Brazovskii-transition at $H=0$, i.e., a fluctuation-induced first order transition in MnSi. 

We wish to thank T. Adams, M. Halder, M. Janoschek, A. Kusmartseva, and S. Mayr for fruitful discussions and assistance with the experiments. Financial support through DFG TRR80 (From Electronic Correlations to Functionality), DFG FOR960 (Quantum Phase Transitions), DFG SFB608, and ERC AdG (291079, TOPFIT) is gratefully acknowledged. A.B. acknowledges financial support through the TUM graduate school.


\begin{thebibliography}{25}
\bibitem[{tri()}]{tricritical}
\bibinfo{note}{A tricritical point refers to a point where three lines of
  critical transitions meet. In our magnetic phase diagram (Fig.\ref{Figure03})
  there is, however, only a single second-order phase transition line that
  emanates from the tricritical point. The other two lines are not visible in
  the two-dimensional magnetic field versus temperature phase diagram. They
  only emerge in a higher dimensional phase diagram that includes a field that
  couples linearly to the helimagnetic order parameter.}
\end{thebibliography}


%

\newpage
\newpage

\appendix

 \setcounter{figure}{0}

\section{Supplementary Material for: Specific heat of the skyrmion lattice phase and field-induced tricritical point in MnSi}

%
%


{\it In this supplement we present a detailed account of the method used to analyze our specific heat data. This concerns in particular the background we subtracted to determine the entropy released at the phase transitions and the change of its temperature dependence with increasing field.}

\vspace{.5em}

In the main text we present specific heat measurements of the magnetic phase diagram of MnSi in the vicinity of the paramagnetic to helimagnetic transition for magnetic fields between zero and $\mu_0 H = 700$ mT, see Fig. 1 of the main text. In order to analyze the evolution of the anomalies at the phase transitions determined at first a magnetic field-dependent background. Our choice of background was defined in the following way as illustrated in Figs.\,1 and 2 
of this supplement. 
We first excluded data points in a temperature range close to the anomalies. The boundaries of this temperature range are illustrated by vertical lines in Figs.\,1 and 2. The remaining data were fitted with a 9$^{\rm th}$ order polynomial which defines a background for a specific choice of boundaries. We considered two extreme choices for these boundaries of excluded data points. On the one hand, the interval was chosen as narrow as possible ensuring that the background still remains smooth, i.e., without generating points of inflection within the background in the temperature range between the two boundaries. In Figs. 1 and 2 this is referred to as high background (green line). On the other hand, the interval was chosen as wide as possible such that the background would still follow the data points as closely up to the anomaly as possible. In Figs. 1 and 2 this is referred to as low background (orange line). As a compromise between these two extremes a middle background has been defined (red line).

In Figs. 1 and 2 data are shown at $H = 0$ and for typical field values in the magnetic phase diagram, notably $\mu_0 H = 180\,{\rm mT}$, 280\,mT and 390\,mT, i.e., a field in the skyrmion lattice phase, a field just above the skyrmion lattice phase and, finally, a field close to the tricritical point, respectively. After subtracting the three different background levels defined above, the resulting anomalies were integrated numerically to determine the change of entropy $\Delta S$ at the transitions. The result is  shown in Fig. 3 of the main text; error bars represent the two limits for the choice of background.

Regardless of the background subtracted from our data the specific heat anomalies display a change of shape as a function of temperature. This is illustrated in Fig. 2, where the anomalies have been fitted by Gaussians (thin blue line), 
\begin{align}
\left.\frac{\Delta C}{T}\right|_{\rm Gauss} = 
	y_0+A e^{- \frac{(T-T_c)^2}{2 w^2}}
\end{align}
The parameters of the fits, $A$, $w$ and $T_c$, are shown in each panel. The offset $y_0$ is below $0.1\,{\rm mJ\,mol^{-1}\,K^2}$ and thus negligible. At zero field and low fields, the anomalies are symmetric and essentially Gaussian; at high fields the shape cannot be fitted by a Gaussian. This is shown for the field $\mu_0 H = 390$ mT close to the tricritical point which clearly demonstrates an incipient asymmetric lambda anomaly characteristic for a second order transition. 

\begin{widetext}

\begin{figure*}[b]
\includegraphics[width=1.0\linewidth,clip=]{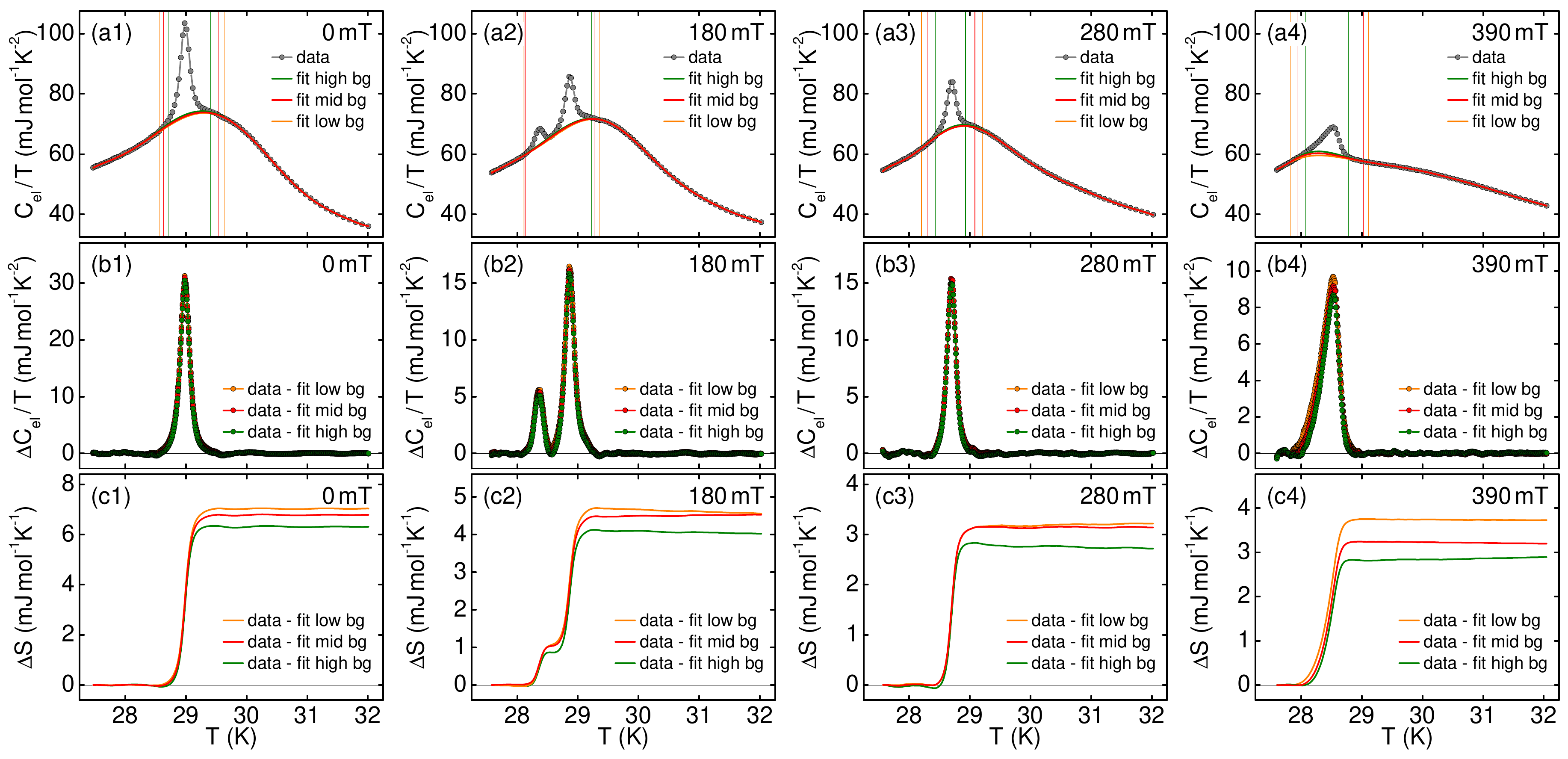}
\caption{Typical specific heat anomalies and step-by-step analysis to determine the change of entropy at the anomalies as shown in Fig. 3 of the main text. For close-up views see Fig. 2 in this supplement. Columns correspond to the same field value. The top row shows the data as measured after subtracting a phonon contribution; colored vertical lines indicate the temperature interval for which data was excluded when determining the background (red line). The centre row shows the specific heat anomalies after subtracting different choices of background as explained in the text. The bottom row displays the entropy as inferred from the specific heat anomaly by numerical integration. Data points in Fig.\,3 of the main text were determined from the non-zero plateau values; the variation of these values with the choice of background defines the error bars in Fig.\,3 of the main text. Panels (a1) through (c1): behavior at zero field. Panels (a2) through (c2): behavior at $\mu_0 H= 180$ mT, i.e., a field value in the skyrmion lattice phase. Panels (a3) through (c3): behavior at $\mu_0 H= 280$ mT, i.e., a field value just above the field range of the skyrmion lattice phase but below the tricritical point. Panels (a4) through (c4): behavior at $\mu_0 H= 390$, i.e.,  close to the tricritical point.
}
\label{figureS1}
\end{figure*}

\begin{figure*}
\includegraphics[width=1.0\linewidth,clip=]{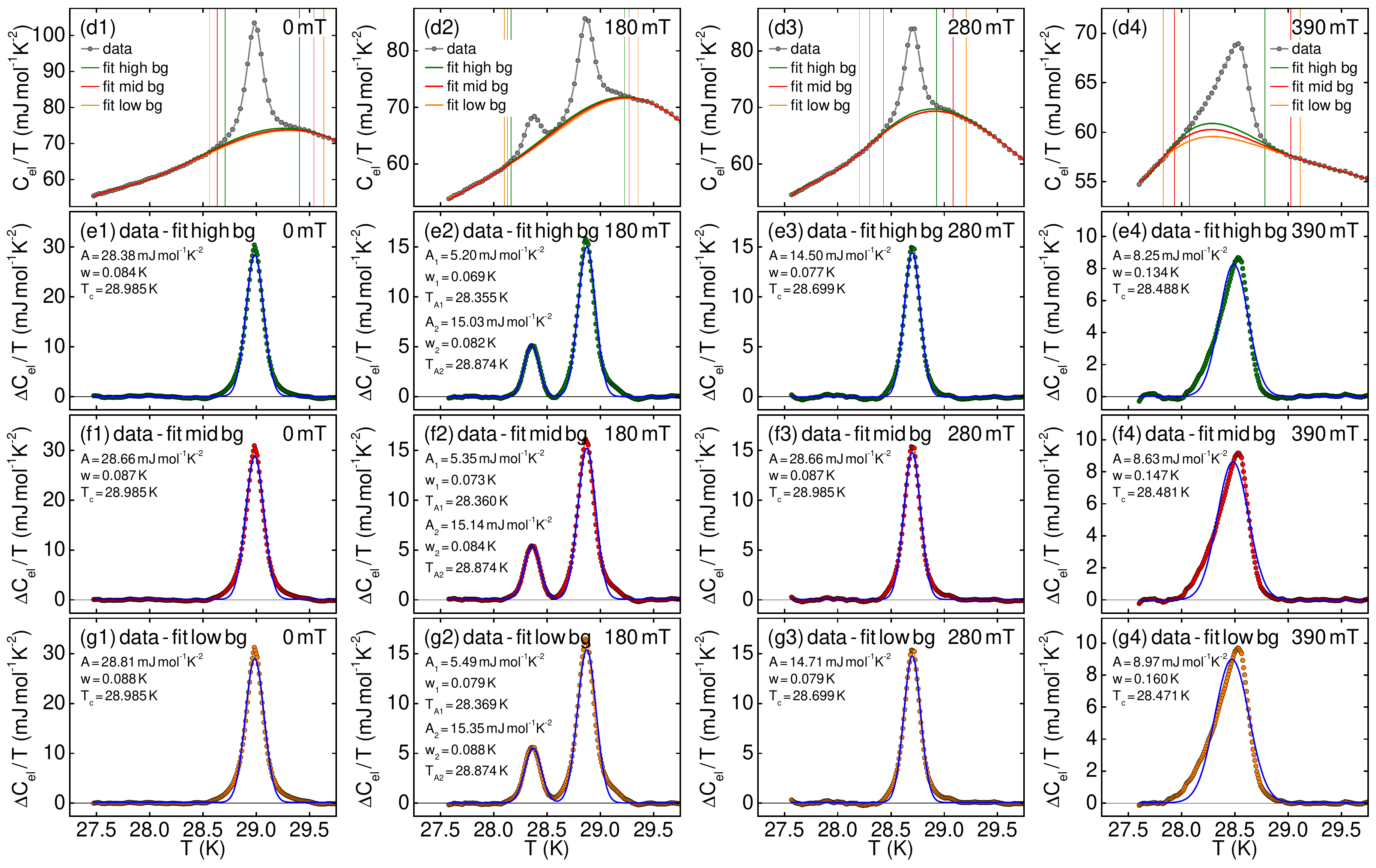}
\caption{Close-up view of the data shown in Fig.\,1 of this supplement. The specific heat anomalies at low fields are well described by a Gaussian but turn asymmetric at high fields indicating the presence of a tricritical point  around $\mu_0 H \gtrsim 390\,{\rm mT}$. Columns correspond to the same magnetic field value. The first row shows the data as measured after subtracting  a phonon contribution. The second, third, and fourth row show the specific heat anomaly after subtracting the high, middle and low background, respectively. The thin blue lines represent Gaussian fits to guide the eye. The shape of the anomaly changes with increasing field from a symmetric, essentially Gaussian shape to an asymmetric shape, which clearly cannot be fitted by a Gaussian. This conclusion is independent of the background subtracted.
}
\label{figureS2}
\end{figure*}

\end{widetext}

\end{document}